# Guided mode meta-optics: Metasurface-dressed nanophotonic waveguides for arbitrary designer mode couplers and on-chip OAM emitters with configurable topological charge

**Yuan Meng[1,3], Tiantian He[1,3], Zhoutian Liu[1], Futai Hu[1], Qirong Xiao[1,2], Qiang Liu[1,2], and Mali Gong[1,2]**

[1] *State Key Laboratory of Precision Measurement Technology and Instruments, Tsinghua University, Beijing 100084, China*
[2] *Key Laboratory of Photonic Control Technology, Ministry of Education, Tsinghua University, Beijing 100084, China*
[3] *Equal contribution: E-mail address: mengy16@mails.tsinghua.edu.cn (Y, Meng) xiaoqirong@mail.tsinghua.edu.cn (Q. Xiao)*

**Abstract:** Metasurfaces have achieved fruitful results in tailoring complexing light fields in free space. However, a systematic investigation on applying the concept of meta-optics to completely control waveguide modes is still elusive. Here we present a comprehensive catalog capable of selectively and exclusively excite almost arbitrary high-order waveguide modes of interest, leveraging silicon metasurface-patterned silicon nitride waveguides. By simultaneously engineering the phase-matched gradient of the metasurface and the vectorial spatial modal overlap between the nanoantenna near-field and target waveguide mode for excitation, either single or multiple high-order modes are successfully launched with high purity reaching 98% and broad bandwidth. Moreover, on-chip twisted light generators are also theoretically demonstrated with configurable OAM topological charge $\ell$ from -3 to +2, serving as a comprehensive framework for metasurface-enabled guided mode optics and motivating further applications such as versatile integrated couplers, demultiplexers, and mode-division multiplexing-based communication systems.

As the electrical bottleneck in conventional electronic circuits are increasingly manifesting in recent years [1], photonic integrated circuits are hatching as promising technologies to potentially revolutionize conventional integrated circuits [2], by providing broadband optical communication systems [3, 4] and ultrafast chip-scale information processing paradigms and optical interconnects with low power consumption [5-7]. However, current integrated photonics encounter limitations from its fundamental building blocks of optical waveguides, in terms of bulk footprint and restrained functionalities [2].

In recent years, there have been growing interest in integrating various subwavelength meta-structures with waveguide platforms [8-18] to enrich the overall structural design library of photonic integrated devices, opening new opportunities to develop novel on-chip photonic devices with either largely enhanced performance or previously hardly accessible sophisticated functionalities [8]. Endeavors are pursued to apply plasmonic nanoantennas on top of dielectric waveguides for off-chip beam manipulations [9] and high-speed polarization demultiplexing [10, 11]. However, there devices inevitably inherit the high Ohmic loss from metal nanoantennas. Dielectric metasurfaces-addressed photonic waveguides are later explored for low-loss devices, where geometric phase (or Pancharatnam-Berry phase) [18] metasurfaces are integrated on silicon [14] and silicon nitride waveguides [15] for compact integrated polarization demultiplexers. Nevertheless, the optical fields that get coupled into these waveguides are uncontrollable hybrid modes [14, 15], which will severely influence their performance in high-speed optical communication systems due to the inter-mode dispersions. Complex light field generations and manipulations via metasurfaces in free-space have been intensively investigated [19, 20]. However, a systematic physical catalog exploring the excellent capability of using metasurface to completely control diverse waveguide modes is still elusive to the best of our knowledge.

In this letter, we present a comprehensive framework targeting the arbitrary manipulation of guided waveguide modes using silicon metasurface-patterned silicon nitride nanophotonic waveguides. Either single or multiple arbitrary high-order modes of interest can be selectively and exclusively excited with high mode purity reaching 98%, by following our proposed easy-to-implement methods to simultaneously engineer the phase gradient of the metasurface and the spatial modal overlap between antenna near fields and the target waveguide mode to excite. Furthermore, via judiciously mixing specific high-order modes, on-chip vortex beam emitters are also theoretically demonstrated with configurable topological charge $\ell$ of orbital angular momentum (OAM) from -3 to +2. This letter can serve as a positive paradigm to migrate meta-optics from free space optics into guided mode optical physics, for catalyzing further researches in photonic integrated circuits and enabling applications such as versatile on-chip couplers, demultiplexers, and OAM-multiplexing-based communication systems [21-24].

The general device structure of the mode-configurable directional coupler is schematically shown in Fig. 1 (a). The normally incident light beam will undergo consecutive scattering events and picks up an effective unidirectional momentum [12, 13] provided by the phase-gradient metasurface atop of the waveguide. To directionally and selectively excite a specific waveguide mode of interest, we first need to match the phase metasurface's phase gradient $\Delta\varphi/d$ with the effective mode index $n_{\mathrm{eff}}$ of the target mode for excitation,

$$(n_t \sin\theta_t - n_i \sin\theta_i)k_0 = n_{\mathrm{eff}}k_0 = \frac{\Delta\varphi}{d}$$

where the $n_i$ and $n_t$ are the material refractive indices of the incident and transmittance medium respectively, incident angle $\theta_i = 0$ under normal incidence, and we have $n_t \sin\theta_t \equiv n_{\mathrm{eff}}$ for guided waves [15, 16]. $k_0 = \sigma \cdot 2\pi/\lambda$ is wavevector, where $|\sigma| = 1$ and $\sigma = \mathrm{sign}(\Delta\varphi/d)$ will determine the propagation direction of the coupled modes.

In addition to the matched phase gradient, judicious engineering of the vectorial spatial modal overlap $\eta$ is also vital for exclusively launching arbitrary high-order waveguide mode (see Figs. 1b and 1c),

$$\eta \propto \frac{|\iint \mathbf{E}_{\mathrm{antenna}}(x,y,z) \cdot \mathbf{E}^{*}_{\mathrm{mode}}(x,y,z)\mathrm{d}y\mathrm{d}z|^2}{(\iint |\mathbf{E}_{\mathrm{antenna}}(x,y,z)|^2\mathrm{d}y\mathrm{d}z) \cdot (\iint |\mathbf{E}_{\mathrm{mode}}(x,y,z)|^2\mathrm{d}y\mathrm{d}z)}$$

where we assume the waveguide is placed along $x$ direction. $\mathbf{E}_{\mathrm{antenna}}$ and $\mathbf{E}_{\mathrm{mode}}$ denote the antenna scattering near-field and the target waveguide mode profile respectively.

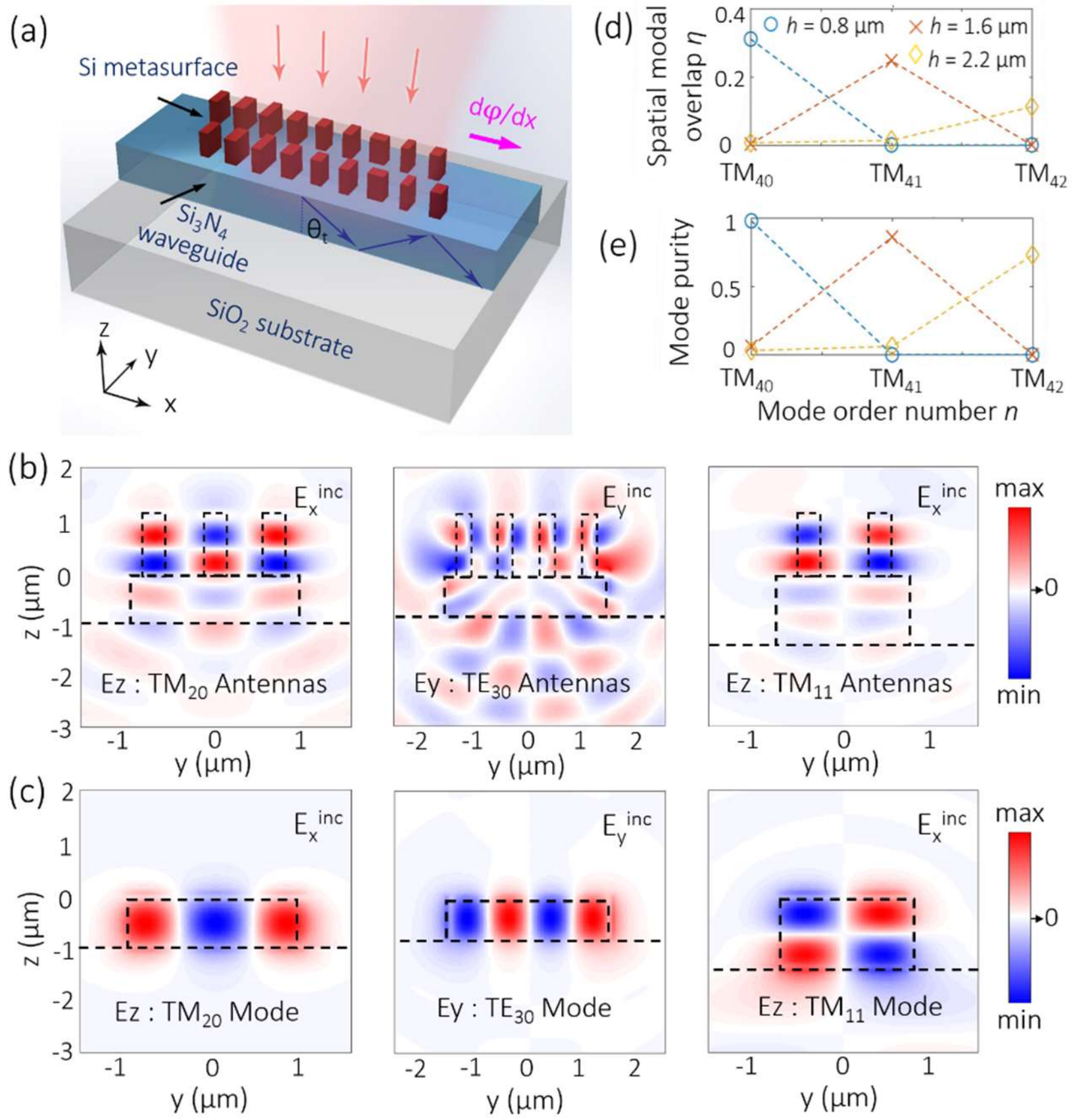

Figure 1. Schematics and spatial modal overlap engineering illustration. (a) Device schematic. Normally incident light can be directionally coupled into a specific guided mode by interacting with the gradient metasurface. (b) Antenna near field. Black dashed lines indicate the profiles of waveguide and the antennas. (c) Electric field distributions of the ideal guided modes. (d) Calculated spatial modal overlap $\eta$ between antenna near field Eantenna and target guided mode $\mathbf{E}_{\mathrm{mode}}$ of devices exclusively launching $TM_{40}$, $TM_{41}$, and $TM_{42}$ modes with same waveguide width of 5.0 μm. (e) Calculated mode purity for the same devices to excite $TM_{40}$, $TM_{41}$, and $TM_{42}$ modes, showing excellent agreement with $\eta$ to validate our assumption.

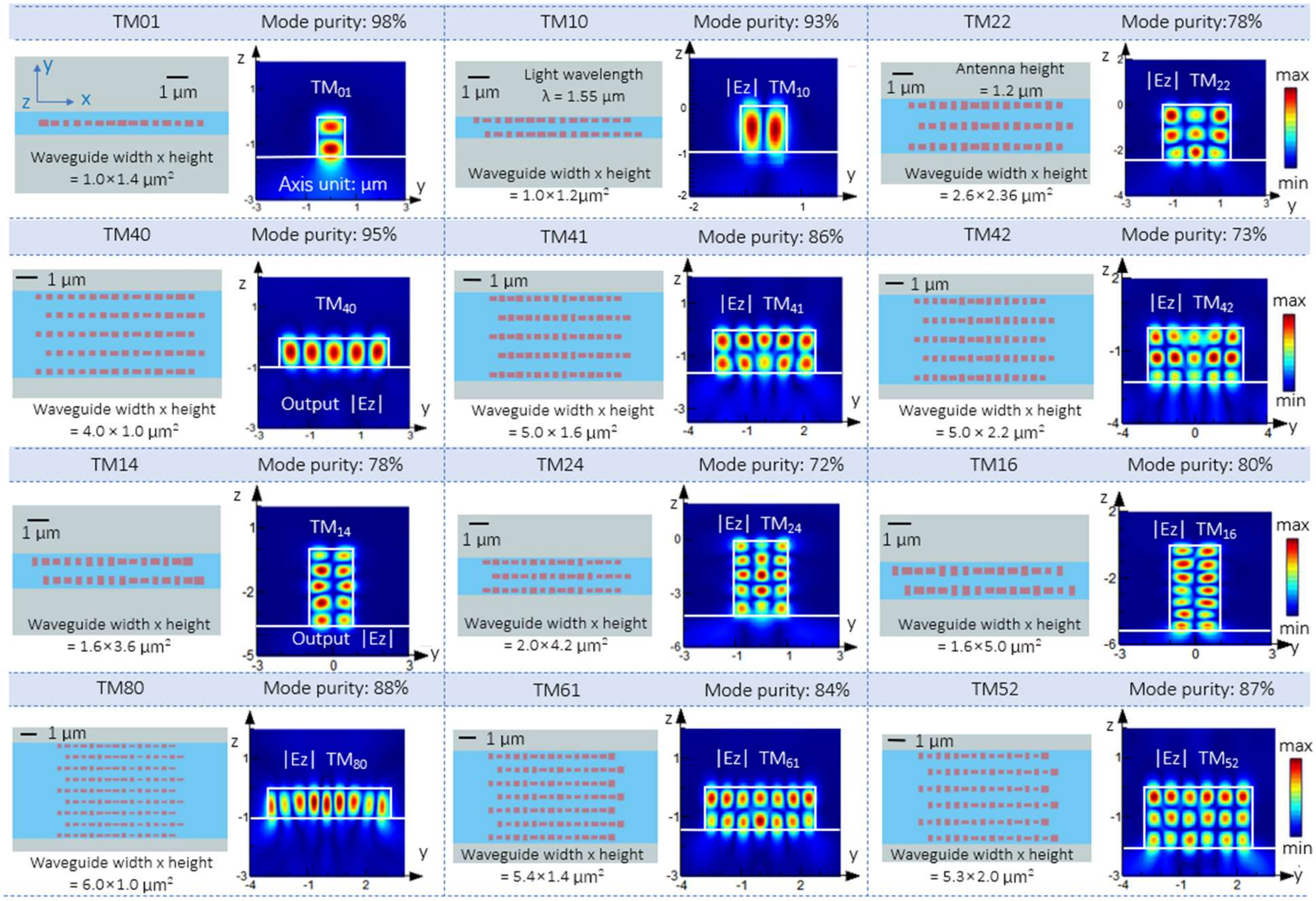

Figure 2. Mode configurable meta-waveguide couplers using silicon metasurface-patterned silicon nitride waveguides to selective launch diverse TM modes. The device schematics (top view) are shown as the left panels with different scale bars, where the silicon nanoantennas are marked as red rectangles; the silicon nitride waveguide is colored in blue; grey color indicates silicon dioxide substrate. The output electric field norm distributions $|\mathbf{E}_z|$ at the right ports are shown as the right panels with mode purity values manifested above. (Structure details are appended in Table 1)

It is worth pointing out that $\eta$ is a vectorial integration. Therefore, the polarization state of $\mathbf{E}_{\text{antenna}}$ and the incident light source is crucial to determine where TE or TM modes are excited. In summary, linearly $x$-polarized light should be generally adopted to excite TM mode series, while $y$-polarizations will launch TE mode series. To selectively excite $\Psi_{m,n}$ mode, $m+1$ rows of identical antenna arrays should be applied with a dislocation $\Delta x$ along $x$ axis between adjacent arrays. The mode order number $n$ is thus controlled by the matched phase gradient $\Delta\varphi/d$ to the effective index $n_{\text{eff}}$ of the target waveguide mode to excite.

We note that in previous similar publications, mode-management is either based on conversion [19] or restrained to only fundamental modes [16]. In contrast, the proposed method in this letter can simultaneously achieve light coupling and mode conversion in a fully integrated and compact manner. Compared with previous research that mostly focuses on versatile polarization and wavelength demultiplexing [17] and only very briefly mentioned high-order mode excitation, this work present a complete framework on deploying meta-optics to tailor guided mode optics, where a systematic investigation on diverse high order waveguide modes with much improved mode purity. Moreover, for the first time, we further extend our approach to launch OAM beams with scalable topological charge to even $\ell = \pm 4$, which is also an obvious increment (approximate one order of magnitude higher) to other similar archived references.

Figure 2 catalogs diverse meta-waveguide couplers to exclusively excite various TE modes as a proof-of-concept, using silicon metasurface addressed silicon nitride photonic waveguides around the telecommunication wavelength of $\lambda = 1.55$ µm. For instance, to selectively launch $\text{TM}_{m,n} = \text{TM}_{5,2}$ modes, $m+1=6$ rows of antennas with matched gradient $\Delta\varphi/d$ are optimized under the illumination of linear $x$-polarized plane wave. Most devices exhibit high mode purity over 80%, and the highest value has reached 98%. Massive TM modes can be also effectively launched in a similar manner as Fig. 3 with averaged mode purity about 90%, by following our integrated design portfolio of guided mode meta-optics.

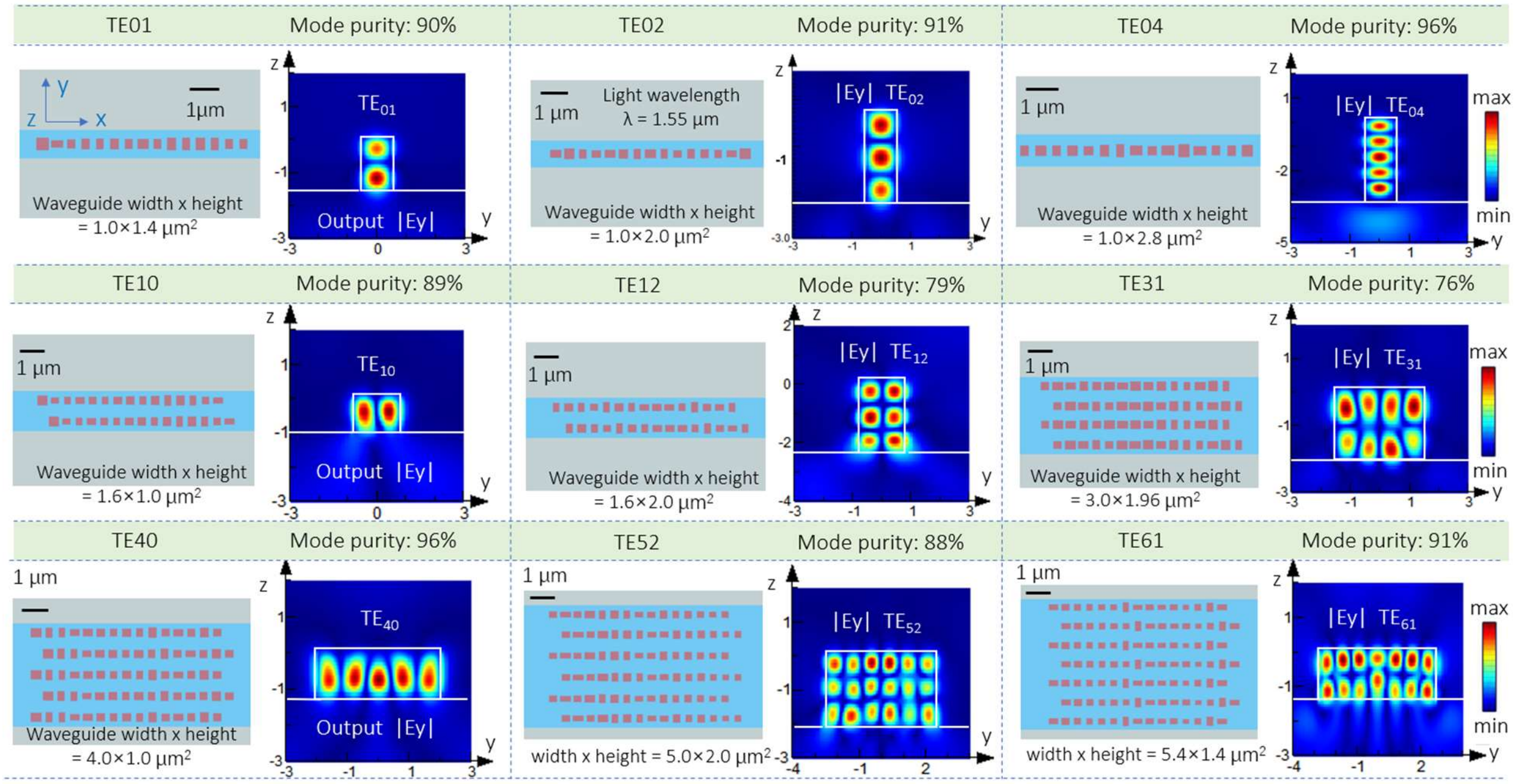

Figure 3. Chip-integrated waveguide mode convertors for exclusively exciting arbitrary TE modes. Device structures and output electric field component distributions $|\mathbf{E}_y|$ are shown in each unit accordingly. (See Table 2 for details)

A practical tip towards high mode purity in designing these meta-devices integrated with phase gradient metasurfaces for given mode $\Psi_{m,p}$ is to judiciously vary the waveguide dimension. If two modes $\Psi_{m,p}$ and $\Psi'_{m,q}$ share same mode order number $m$ (denoting same rows of antenna arrays are involved), and they by accident have approaching value of effective mode index $n_{\text{eff}}$, crosstalk will take place. In this case, it will be easy to simultaneously excite both modes $\Psi_{m,p}$ and $\Psi'_{m,q}$, which will deteriorate the mode purity of the interested mode $\Psi_{m,p}$. An easy method to suppress the unwanted $\Psi'_{m,q}$ mode is to modify the dimension of the waveguide to discriminate and diverge the effective index $n_{\text{eff}}$ of the two modes. Then the mode purity of the target $\Psi_{m,p}$ mode will be effectively enhanced.

Beside eigen-TE or TM mode series, our proposed method can also facilitate manipulation complex intra-waveguide light fields. Selective excitation of either single or multiple specific waveguide modes (eigen- or certain hybrid modes) is also available. Taking optical vortex beams carrying OAM as an instance, complex OAM field with scalable and configurable topological charge $\ell$ can be realized using our proposed method and meta-waveguide platform.

An easy approach is to utilize mode mixing method [23], where specific OAM light can be obtained by wisely mixing several high-order modes. Leveraging the abovementioned catalog for selective launching specific high-order modes of interest, we have numerically validated on-chip OAM generations with configurable topological charge $\ell$ from -3 to +3 using FDTD calculations. Fig. 4 shows some preliminary results for integrated twisted light generators for $OAM_{-2}$, $OAM_{-1}$ and $OAM_{+1}$, which are realized by properly mixing specific high order modes. This mechanism is schematically shown as the right panels, where (taking the first row for $OAM_{-2}$ as an instance) the mode distributions are shown when only one of the antenna arrays (i.e. namely $TE_{20}$ antennas, $TE_{02}$ antennas or $TE_{11}$ antennas) are present. The phase distributions of the mixed output mode at right waveguide ports are also shown, indicating designer topological charges $\ell = -2, +1,$ and $-1$ respectively.

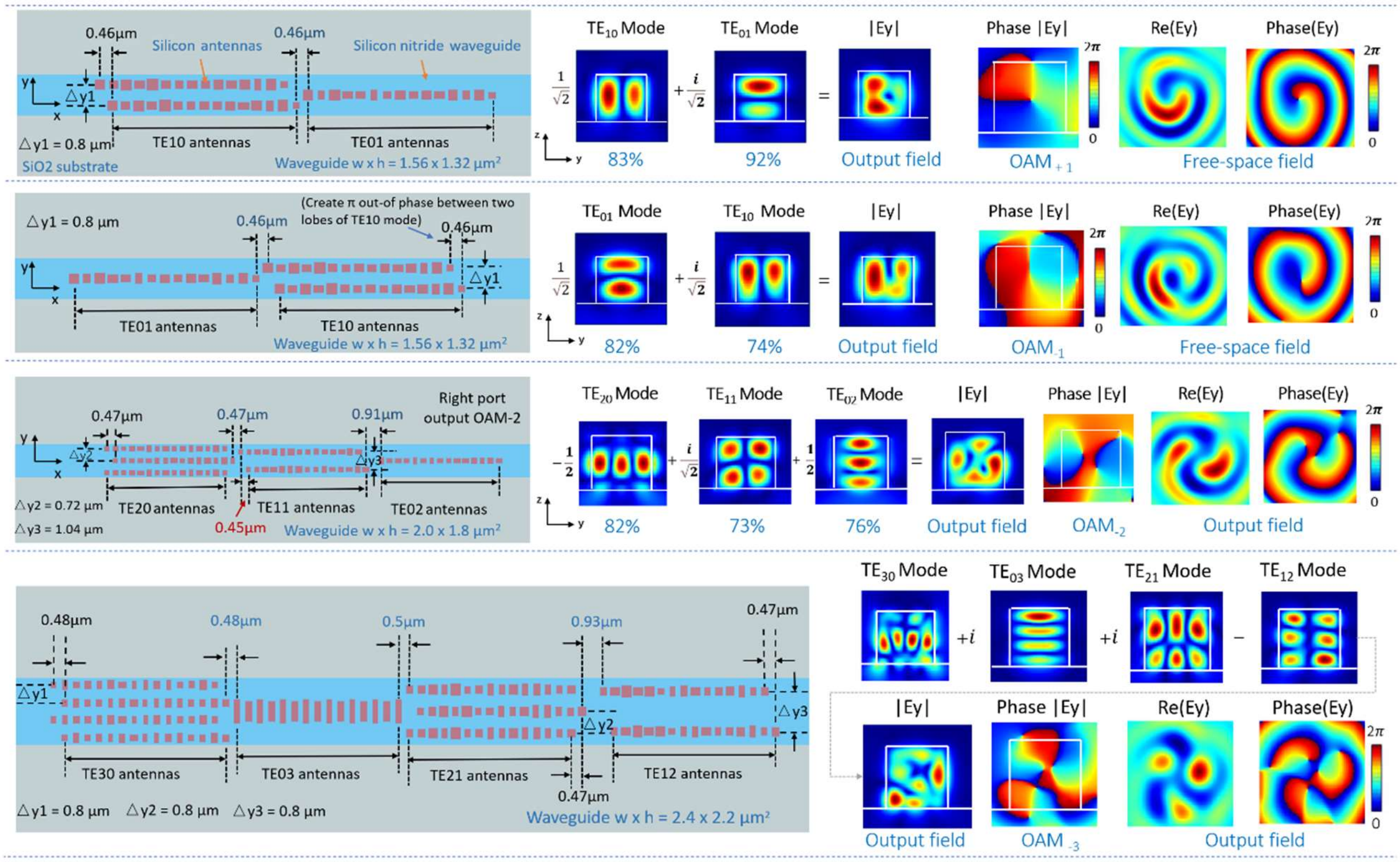

Figure 4. On-chip vortex beam generators with configurable OAM topological charges for (a) $\ell = +1$, (b) $\ell = -1$, (c) $\ell = -2$, and (d) $\ell = -3$. The device schematics (top view) are shown as left panels. The output electric field distributions for individual antenna arrays, the overall OAM field inside the meta-waveguide, and the output optical fields after exiting waveguide Right Ports are shown as right panels. (See Tables 3-6 for device structure details)

We further note that this scenario is also potentially scalable for higher OAM modes with simpler device structure than those exploring Parity-Time symmetry. Moreover, as shown in Fig. 5, the devices to produce light vortices can have multiple operation wavelengths and broader bandwidth if compared with the micro-ring resonators approach.

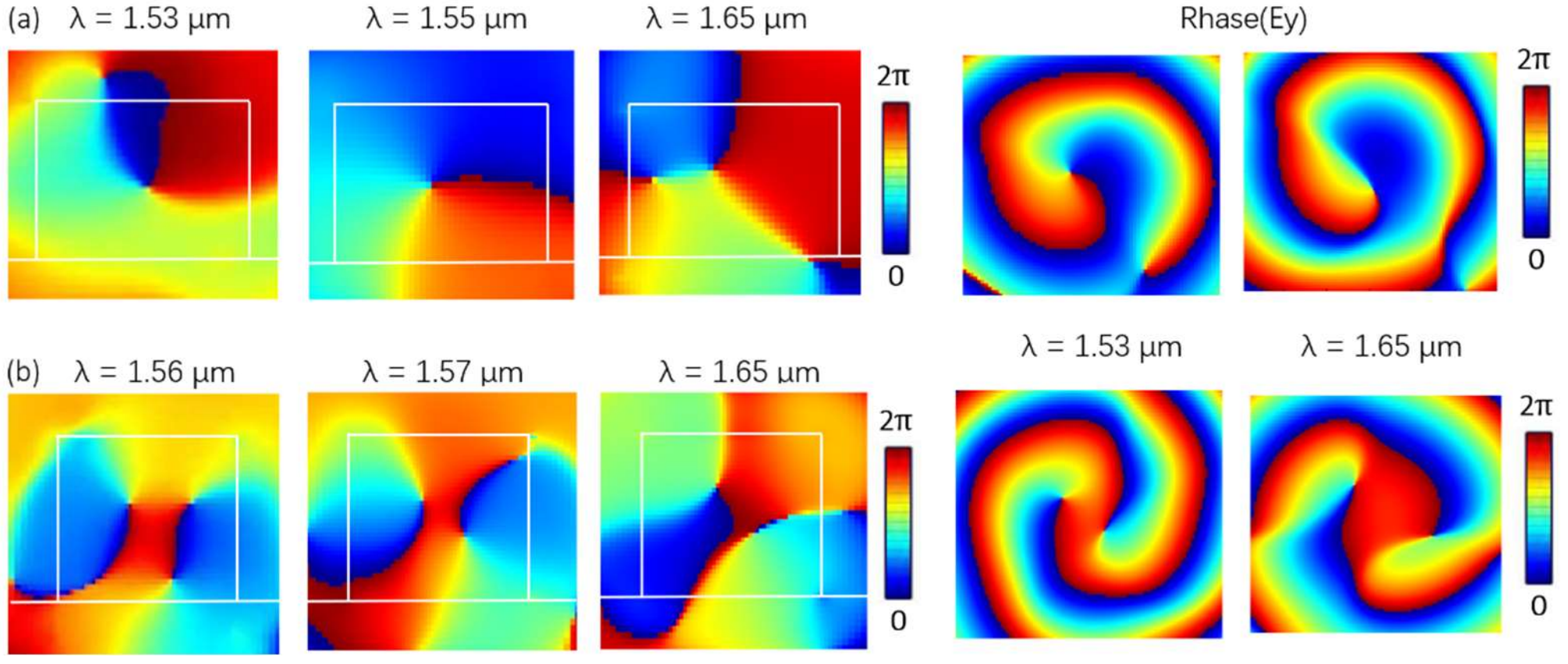

Figure 5. Phase distribution of $OAM_{+1}$ and $OAM_{-2}$ modes at different wavelengths while inside (left panels) and outside the waveguides propagating in free space (right panels) showing multiple working wavelengths.

In summary, we have extended the concept of meta-optics into the realm of integrated optics to achieve almost arbitrary control over waveguide modes. By simultaneously engineering the phase gradient of the silicon metasurface resting on the silicon nitride waveguide and the vectorial spatial modal overlap between antenna near field and target waveguide mode, we can selectively and exclusively excite almost arbitrary high-order modes of interest with high mode purity reaching 98%. By judiciously mixing several high-order modes, structured light like optical vortices carrying OAM can be also excited on a photonic chip with configurable topological charge from -3 to +2. The proposed device may be further optimized via inverse design algorithms [25, 26]. The meta-waveguides can be also allied with two-dimensional materials [27-31], phase-change materials [32-34] or lithium niobate [35-37] for reconfigurable or dynamically tunable devices. Through further optimizations, we believe OAM beams with even higher topological charge is also possible, opening new opportunities for chip-scale structured light generations and potentially boosted mode-division-multiplexing-based communication systems.

## Appendix: Device structure parameters and details.

Table 1: Detailed structure parameters for devices to excite various TM modes in Figure 1.

| $m$ | -7 | -6 | -5 | -4 | -3 | -2 | -1 | 0 | 1 | 2 | 3 | 4 | 5 | 6 | 7 | $\Delta x$/μm | Y/μm |
|---|---|---|---|---|---|---|---|---|---|---|---|---|---|---|---|---|---|
| lx($TM_{01}$) | 336 | 420 | 292 | 324 | 404 | 444 | 272 | 432 | 332 | 336 | 272 | 412 | 448 | 372 | 440(nm) | 0.48 | Y=0 |
| ly($TM_{01}$) | 296 | 264 | 312 | 344 | 296 | 324 | 352 | 336 | 308 | 408 | 368 | 368 | 300 | 440 | 304(nm) | | |
| lx($TM_{10}$) | 264 | 412 | 264 | 348 | 428 | 388 | 412 | 312 | 260 | 376 | 384 | 320 | 320 | 356 | 348(nm) | 0.52 | $Y_{up}$=0.33 |
| ly($TM_{10}$) | 292 | 216 | 280 | 268 | 224 | 264 | 224 | 284 | 280 | 244 | 224 | 216 | 228 | 212 | 224(nm) | | |
| lx($TM_{22}$) | 296 | 376 | 260 | 368 | 260 | 320 | 436 | 392 | 272 | 340 | 300 | 416 | 296 | 328 | 304(nm) | 0.49 | $Y_{up}$=0.98 |
| ly($TM_{22}$) | 316 | 276 | 400 | 332 | 408 | 404 | 296 | 376 | 356 | 384 | 320 | 280 | 324 | 416 | 320(nm) | | $Y_{mid}$=0 |
| lx($TM_{40}$) | 264 | 288 | 268 | 300 | 288 | 204 | 400 | 332 | 264 | 404 | 268 | 384 | 332 | 436 | 280(nm) | 0.47 | $Y_{up1}$=1.72 |
| ly($TM_{40}$) | 248 | 284 | 248 | 256 | 236 | 268 | 224 | 272 | 276 | 216 | 272 | 288 | 232 | 264 | 260(nm) | | $Y_{up2}$=0.87 |
| lx($TM_{41}$) | 440 | 344 | 440 | 404 | 284 | 400 | 304 | 332 | 260 | 440 | 312 | 396 | 384 | 272 | 436(nm) | 0.50 | $Y_{up1}$=2.14 |
| ly($TM_{41}$) | 296 | 380 | 296 | 348 | 332 | 340 | 316 | 376 | 420 | 276 | 312 | 340 | 300 | 304 | 296(nm) | | $Y_{up2}$=1.06 |
| lx($TM_{42}$) | 328 | 312 | 268 | 356 | 428 | 304 | 384 | 444 | 260 | 332 | 260 | 356 | 268 | 368 | 320(nm) | 0.49 | $Y_{up1}$=2.14 |
| ly($TM_{42}$) | 304 | 364 | 360 | 340 | 296 | 432 | 300 | 332 | 428 | 396 | 448 | 352 | 376 | 360 | 312(nm) | | $Y_{up2}$=1.06 |
| lx($TM_{14}$) | 260 | 352 | 344 | 292 | 268 | 292 | 256 | 288 | 400 | 384 | 260 | 420 | 252 | 436 | 448(nm) | 0.51 | $Y_{up}$=0.43 |
| ly($TM_{14}$) | 392 | 332 | 300 | 376 | 348 | 420 | 396 | 416 | 292 | 256 | 360 | 312 | 404 | 308 | 368(nm) | | |
| lx($TM_{24}$) | 432 | 364 | 352 | 404 | 272 | 376 | 280 | 328 | 448 | 384 | 256 | 324 | 448 | 340 | 344(nm) | 0.52 | $Y_{up}$=0.75 |
| ly($TM_{24}$) | 252 | 312 | 256 | 232 | 312 | 332 | 288 | 312 | 244 | 316 | 376 | 228 | 240 | 220 | 244(nm) | | $Y_{mid}$=0 |
| lx($TM_{16}$) | 260 | 444 | 376 | 412 | 256 | 308 | 260 | 396 | 440 | 300 | 428 | 292 | 404 | 264 | 252(nm) | 0.52 | $Y_{up}$=0.42 |
| ly($TM_{16}$) | 432 | 332 | 300 | 276 | 448 | 412 | 392 | 272 | 292 | 360 | 292 | 416 | 292 | 284 | 420(nm) | | |
| lx($TM_{80}$) | 264 | 368 | 288 | 388 | 260 | 412 | 408 | 440 | 260 | 316 | 260 | 316 | 340 | 408 | 436(nm) | 0.51 | $Y_{up1}$=2.8 $Y_{up2}$=2.12 |
| ly($TM_{80}$) | 256 | 212 | 180 | 208 | 284 | 192 | 148 | 188 | 264 | 136 | 256 | 212 | 152 | 240 | 144(nm) | | $Y_{up3}$=1.4 $Y_{up4}$=0.7 |
| lx($TM_{61}$) | 268 | 320 | 276 | 336 | 260 | 440 | 272 | 384 | 288 | 372 | 264 | 412 | 372 | 408 | 388(nm) | 0.53 | $Y_{up1}$=2.36 $Y_{up2}$=1.58 |
| ly($TM_{61}$) | 344 | 240 | 300 | 328 | 368 | 224 | 296 | 320 | 264 | 216 | 300 | 264 | 240 | 260 | 396(nm) | | $Y_{up3}$=0.78 |
| lx($TM_{52}$) | 264 | 280 | 260 | 312 | 264 | 428 | 276 | 308 | 300 | 352 | 256 | 448 | 252 | 280 | 424(nm) | 0.53 | $Y_{up1}$=2.33 $Y_{up2}$=1.42 |
| ly($TM_{52}$) | 364 | 252 | 392 | 372 | 332 | 224 | 288 | 352 | 256 | 220 | 344 | 204 | 364 | 204 | 388(nm) | | $Y_{up1}$=0.46 |

Table 2: Detailed structure parameters for devices in Figure 2 to launch diverse TE modes.

| $m$ | -7 | -6 | -5 | -4 | -3 | -2 | -1 | 0 | 1 | 2 | 3 | 4 | 5 | 6 | 7 | $\Delta x$/μm | Y/μm |
|---|---|---|---|---|---|---|---|---|---|---|---|---|---|---|---|---|---|
| lx($TE_{01}$) | 444 | 432 | 288 | 292 | 260 | 308 | 280 | 392 | 272 | 320 | 264 | 340 | 268 | 260 | 284(nm) | 0.47 | Y = 0 |
| ly($TE_{01}$) | 440 | 260 | 312 | 380 | 392 | 372 | 336 | 344 | 360 | 436 | 448 | 428 | 444 | 340 | 356(nm) | | |
| lx($TE_{02}$) | 404 | 388 | 272 | 344 | 396 | 332 | 260 | 404 | 260 | 296 | 260 | 308 | 272 | 396 | 408(nm) | 0.53 | Y = 0 |
| ly($TE_{02}$) | 296 | 396 | 344 | 280 | 292 | 352 | 384 | 328 | 368 | 280 | 356 | 328 | 316 | 256 | 388(nm) | | |
| lx($TE_{04}$) | 272 | 348 | 272 | 340 | 316 | 292 | 252 | 384 | 288 | 404 | 340 | 400 | 272 | 272 | 348(nm) | 0.54 | Y = 0 |
| ly($TE_{04}$) | 340 | 332 | 332 | 328 | 296 | 376 | 416 | 308 | 296 | 300 | 416 | 296 | 300 | 340 | 260(nm) | | |
| lx($TE_{10}$) | 400 | 320 | 268 | 368 | 308 | 260 | 356 | 336 | 272 | 336 | 264 | 332 | 268 | 296 | 396(nm) | 0.46 | $Y_{up}$=0.42 |
| ly($TE_{10}$) | 408 | 252 | 320 | 264 | 296 | 300 | 292 | 336 | 336 | 368 | 416 | 408 | 388 | 292 | 252(nm) | | |
| lx($TE_{12}$) | 264 | 364 | 264 | 312 | 256 | 356 | 264 | 384 | 380 | 388 | 276 | 276 | 324 | 368 | 256(nm) | 0.52 | $Y_{up}$=0.42 |
| ly($TE_{12}$) | 364 | 336 | 356 | 280 | 412 | 320 | 344 | 268 | 288 | 264 | 308 | 412 | 288 | 252 | 352(nm) | | |
| lx($TE_{31}$) | 332 | 380 | 396 | 324 | 284 | 372 | 300 | 352 | 260 | 380 | 260 | 312 | 280 | 324 | 416(nm) | 0.48 | $Y_{up1}$=1.14 |
| ly($TE_{31}$) | 292 | 316 | 288 | 276 | 312 | 324 | 304 | 324 | 388 | 352 | 404 | 400 | 324 | 396 | 292(nm) | | $Y_{up2}$=0.36 |
| lx($TE_{40}$) | 412 | 268 | 252 | 356 | 372 | 328 | 284 | 356 | 268 | 304 | 316 | 396 | 288 | 336 | 320(nm) | 0.47 | $Y_{up1}$=1.64 |
| ly($TE_{40}$) | 352 | 388 | 376 | 252 | 288 | 316 | 308 | 316 | 336 | 392 | 300 | 276 | 316 | 376 | 304(nm) | | $Y_{up2}$=0.82 |
| lx($TE_{52}$) | 328 | 444 | 320 | 432 | 264 | 340 | 264 | 348 | 348 | 304 | 268 | 296 | 280 | 272 | 276(nm) | 0.52 | $Y_{up1}$=2.12, $Y_{up1}$=1.31 |
| ly($TE_{52}$) | 264 | 232 | 264 | 300 | 356 | 340 | 340 | 232 | 252 | 340 | 3p4 | 32p | 276 | 236 | 272(nm) | | $Y_{up3}$=0.44 |
| lx($TE_{61}$) | 392 | 344 | 288 | 252 | 300 | 272 | 252 | 384 | 340 | 332 | 332 | 272 | 276 | 272 | 404(nm) | 0.52 | $Y_{up1}$=2.36, $Y_{up2}$=1.58 |
| ly ($TE_{61}$) | 248 | 288 | 272 | 268 | 256 | 240 | 392 | 212 | 244 | 208 | 244 | 200 | 264 | 380 | 236(nm) | | $Y_{up1}$=0.78 |

(The following Tables manifest the structure parameters for Fig. 3)

Table 3: Device parameters for the on-chip $OAM_{+1}$ mode generator with topological charge $\ell = +1$.

| $m$ | -7 | -6 | -5 | -4 | -3 | -2 | -1 | 0 | 1 | 2 | 3 | 4 | 5 | 6 | 7 | Antenna center/μm coordinate | Y/μm |
|---|---|---|---|---|---|---|---|---|---|---|---|---|---|---|---|---|---|
| lx($TE_{10}$) | 388 | 352 | 416 | 392 | 440 | 356 | 312 | 392 | 336 | 376 | 420 | 420 | 268 | 380 | 268(nm) | $X_{up}$(m)=0.5×m | $Y_{up}$=0.4 |
| ly($TE_{10}$) | 392 | 308 | 412 | 312 | 440 | 320 | 304 | 332 | 304 | 356 | 300 | 364 | 420 | 448 | 260(nm) | $X_{mid}$(m)=0.5×m+0.46 | |
| lx($TE_{01}$) | 380 | 276 | 424 | 432 | 388 | 280 | 312 | 392 | 336 | 376 | 420 | 336 | 296 | 256 | 276(nm) | $X_{mid}$(m)=0.5×m+7.92 | Y = 0 |
| ly($TE_{01}$) | 392 | 364 | 420 | 308 | 292 | 412 | 304 | 332 | 304 | 352 | 300 | 436 | 332 | 420 | 256(nm) | | |

Table 4: Device parameters for the on-chip $OAM_{-1}$ mode generator with topological charge $\ell = -1$.

| $m$ | -7 | -6 | -5 | -4 | -3 | -2 | -1 | 0 | 1 | 2 | 3 | 4 | 5 | 6 | 7 | Antenna center/μm coordinate | Y/μm |
|---|---|---|---|---|---|---|---|---|---|---|---|---|---|---|---|---|---|
| lx($TE_{01}$) | 380 | 276 | 424 | 432 | 388 | 280 | 312 | 392 | 336 | 376 | 420 | 336 | 296 | 256 | 276(nm) | $X_{up}$(m)=0.5×m | Y = 0 |
| ly($TE_{01}$) | 392 | 364 | 420 | 308 | 292 | 412 | 304 | 332 | 304 | 352 | 300 | 436 | 332 | 420 | 256(nm) | | |
| lx($TE_{10}$) | 388 | 352 | 416 | 392 | 440 | 356 | 312 | 392 | 336 | 376 | 420 | 420 | 268 | 380 | 268(nm) | $X_{up}$(m)=0.5×m+7.46 | $Y_{up}$=0.4 |
| ly($TE_{10}$) | 392 | 308 | 412 | 312 | 440 | 320 | 304 | 332 | 304 | 356 | 300 | 364 | 420 | 448 | 260(nm) | $X_{low}$(m)=0.5×m+7.91 | |

Table 5: Device parameters for the on-chip $OAM_{-2}$ mode generator with topological charge $\ell = -2$.

| $m$ | -7 | -6 | -5 | -4 | -3 | -2 | -1 | 0 | 1 | 2 | 3 | 4 | 5 | 6 | 7 | Antenna center/μm coordinate | Y/μm |
|---|---|---|---|---|---|---|---|---|---|---|---|---|---|---|---|---|---|
| $lx(TE_{20})$ | 300 | 324 | 320 | 348 | 404 | 320 | 272 | 428 | 300 | 368 | 263 | 432 | 280 | 364 | 280(nm) | $X_{up}(m)=0.5\times m$ | $Y_{up}=0.72$ |
| $ly(TE_{20})$ | 304 | 336 | 300 | 332 | 296 | 384 | 352 | 336 | 320 | 348 | 420 | 356 | 352 | 428 | 360(nm) | $X_{mid}(m)=0.5\times m+0.47$ | $Y_{mid}=0$ |
| $lx(TE_{11})$ | 272 | 272 | 424 | 412 | 352 | 272 | 444 | 292 | 300 | 304 | 376 | 364 | 424 | 340 | 436(nm) | $X_{up}(m)=0.5\times m+7.95$ | $Y_{up}=0.52$ |
| $ly(TE_{11})$ | 304 | 360 | 356 | 304 | 416 | 448 | 376 | 364 | 308 | 372 | 300 | 344 | 300 | 436 | 304(nm) | $X_{low}(m)=0.5\times m+8.40$ | |
| $lx(TE_{02})$ | 360 | 364 | 392 | 364 | 252 | 396 | 264 | 364 | 308 | 420 | 444 | 372 | 280 | 368 | 408(nm) | $X_{up}(m)=0.5\times m+16.32$ | Y=0 |
| $ly(TE_{02})$ | 448 | 312 | 292 | 320 | 436 | 324 | 364 | 276 | 312 | 332 | 296 | 352 | 344 | 396 | 304(nm) | | |

Table 6: Device parameters for the on-chip $OAM_{-3}$ mode generator with topological charge $\ell = -3$.

| $m$ | -7 | -6 | -5 | -4 | -3 | -2 | -1 | 0 | 1 | 2 | 3 | 4 | 5 | 6 | 7 | Antenna center/μm coordinate | Y/μm |
|---|---|---|---|---|---|---|---|---|---|---|---|---|---|---|---|---|---|
| $lx(TE_{30})$ | 252 | 240 | 272 | 288 | 220 | 276 | 352 | 204 | 212 | 304 | 216 | 268 | 268 | 320 | 280(nm) | $X_{up}(m)=0.5\times m$ | $Y_{up1}=0.95$ |
| $ly(TE_{30})$ | 256 | 344 | 256 | 292 | 304 | 332 | 248 | 280 | 360 | 304 | 340 | 240 | 264 | 344 | 360(nm) | $X_{mid}(m)=0.5\times m+0.48$ | $Y_{up2}=0.3$ |
| $lx(TE_{03})$ | 256 | 296 | 256 | 296 | 256 | 296 | 256 | 296 | 256 | 296 | 256 | 296 | 256 | 296 | 256(nm) | $X_{mid}(m)=0.5\times m+7.97$ | Y=0 |
| $ly(TE_{03})$ | 836 | 720 | 836 | 720 | 836 | 720 | 836 | 720 | 836 | 720 | 836 | 720 | 836 | 720 | 836(nm) | | |
| $lx(TE_{21})$ | 312 | 348 | 416 | 284 | 440 | 252 | 360 | 280 | 264 | 400 | 256 | 300 | 272 | 284 | 364(nm) | $X_{up}(m)=0.5\times m+15.47$ | $Y_{up}=0.78$ |
| $ly(TE_{21})$ | 292 | 308 | 392 | 352 | 416 | 268 | 292 | 404 | 360 | 356 | 440 | 440 | 356 | 300 | 304(nm) | $X_{mis}(m)=0.5\times m+15.94$ | $Y_{mid}=0$ |
| $lx(TE_{12})$ | 280 | 440 | 400 | 408 | 448 | 304 | 256 | 280 | 432 | 272 | 328 | 364 | 260 | 292 | 304(nm) | $X_{up}(m)=0.5\times m+23.87$ | $Y_{up}=0.72$ |
| $ly(TE_{12})$ | 304 | 300 | 424 | 308 | 368 | 252 | 384 | 404 | 292 | 296 | 304 | 336 | 428 | 296 | 320(nm) | $X_{mis}(m)=0.5\times m+24.34$ | |